# High second harmonic generation in ferroelectric nematic liquid crystals by doping with optimally oriented chromophores

J. Ortega*[a], C.L. Folcia[a] and T. Sierra[b]

[a]*Department of Physics, Faculty of Science and Technology, UPV/EHU, Bilbao, Spain;*
[b]*Instituto de Nanociencia y Materiales de Aragón (INMA), Química Orgánica, Facultad de Ciencias. CSIC-Universidad de Zaragoza, Pedro Cerbuna 12, ES-50009 Zaragoza, Spain*

e-mail address: josu.ortega@ehu.eus

# High second harmonic generation in ferroelectric nematic liquid crystals by doping with optimally oriented chromophores

We report on the second harmonic generation performance of a ferroelectric nematic fluid consisting of the prototype ferroelectric nematic liquid crystal, RM734 mixed with the classical chromophore Dispersed-Orange 3 at 5 %wt. The mixture exhibits an interesting mesomorphic behavior with a wide temperature range of the ferroelectric mesophase, high nematic order parameter and ease of alignment in cells. In the study, two fundamental wavelengths (1064 nm and 1574 nm) have been used in order to account for the second harmonic generation performance in transparent and absorbing regimens. The results have been very outstanding and are among the highest for liquid crystals so far. Specifically, second order susceptibility tensor components of up to 25 pmV$^{-1}$ and 8.5 pmV$^{-1}$ have been obtained in the absorbing and transparent regimens, respectively. The mixing of chromophores with ferroelectric nematic liquid crystals is then stablished as a promising strategy in the search of high-performance nonlinear-optical mesogens.

## 1. Introduction

The search for new materials with high performance in nonlinear optics (NLO) is a field of intense research activity. Interest focuses on the multiple technological applications based on NLO effects such as optic frequency doublers, Pockels and Kerr cells or ultrafast modulators, among others. In this sense, liquid crystals (LCs) have been considered interesting candidates, and intense research activity has been carried out in this field for decades [1-3]. Recently, the emergence of ferroelectric nematic materials in the LC landscape (FNLCs) has revived interest in LCs as materials for nonlinear optics (NLO). Specifically, very promising results have recently been reported in several scientific works on second order nonlinear effects [4-13]. Apart from the advantages that LCs for NLO generally present, such as the easy and versatility of synthesis, or their fluidity, which facilitates the implementation in electronic devices, FNLCs present a singular symmetry that make them very suitable for NLO purposes. This phase has been recently discovered [14-26] and gathers the main features of the classical nematic phase, i.e., it is a uniaxial fluid in which the molecular constituents only present orientational order along the so-called molecular director but, contrary to the nematic phase, the ferroelectric variant does not present the head-to-tail invariance and, therefore, a net polarization appears along the molecular director. This fact is very interesting to achieve high performance second order NLO effects, since long donor-acceptor groups oriented along the polarization direction can be hosted by the constituent molecules. Therefore, compared to other different LCs materials designed for second order NLO purposes such as bent-core molecules, dimers or trimers [27-30], the FNLCs present an optimal molecular arrangement and, in addition, the molecular size can be smaller. As a consequence, the density of chromophores per unit volume can be higher. One of the most interesting second order effects is the second harmonic generation (SHG). This nonlinear effect is driven by the so called second order susceptibility tensor ($d_{ijk}$) that must be characterized in order to obtain a complete description of the phenomenon. Recently, the characterization of the SHG efficiency of a few FNLC prototypes has been carried out with very promising results [7,9,12]. However, the obtained performances are far from being suitable for applications and it is worth dedicating more investigation in this regard.

Unfortunately, despite the intense research activity in the field of FNLC, the number of compounds synthetized up to know is quite limited [4,6,9,14-16,22,23,25] and the comprehension of the mechanisms that give rise to this phase are not fully understood [4]. This limitation can be even more important if the synthesis is aimed to a given application. In this respect, the possibility of implementing long donor-acceptor groups in the molecular structure, maintaining the ferroelectric nematic character of the phase, has not been addressed in the scientific literature so far.

Recently, the possibility of using blends of FNLC with classical chromophores was suggested [7,31] and several scientific works have demonstrated the suitability of this strategy [9,13]. Xia et al [9] studied the mesomorphic behaviour of blends of the FNLC prototype, DIO with several chromophores at different concentrations. All the mixtures exhibited the FN phase over wide temperature ranges even for concentrations up to 10% wt%. Following this strategy, improvements of more than one order of magnitude were achieved in some of the measured $d_{ijk}$ values. However, the host DIO presents a poor performance ($d_{ijk}$ components up to 0.24 pmV$^{-1}$) and, in addition, the $d_{ijk}$ values are absorption enhanced due to proximity of the absorption bands of the doping chromophores to the second harmonic (SH) light wavelength (532 nm), thus reducing the actual performance of the material.

More recently, Karthick et al [13], have investigated materials made from mixtures of DIO with the TMC chromophore. In this case, all the blends are transparent for the SH light of the experiment (532 nm). They studied chromophore concentrations up to 7 wt%. Similar to the previous case, the mixtures present acceptable temperature ranges of the FN phase with a descending trend when increasing the concentration. In this work, the highest $d_{ijk}$ value achieves 4 pmV$^{-1}$. Although the improvement of the performance in transparent regime is very noticeable, the result is still below the efficiency of pure FNLC, RM734 (d = 5.6 pmV$^{-1}$). Therefore, it is worth inspecting the feasibility of mixing RM734 with high-performance chromophores.

In this work, we report a complete SHG study of a mixture of the FNLC RM734 with the classical chromophore Dispersed-Orange3 at 5 wt%. We have selected this concentration since, according to Xia et al [9] and Karthick et al. [13], the SHG performances of mixtures of FNLCs with chromophores started to decline about this limit. The mixture presents an excellent mesomorphic behaviour with a similar temperature range for the FN phase than the pure RM734. We have performed a study of the SHG performance by using fundamental light of two different wavelengths, 1064 nm and 1574 nm, in order to characterize the NLO efficiency both in absorption and transparent regimes, respectively. The SHG results (25 pmV$^{-1}$ at 1064nm, 8.5 pmV$^{-1}$ at 1574 nm) clearly improve those of the pure host (5.6 pmV$^{-1}$ at 1064 nm, in transparent regime) and are among the best performances in LCs so far.

The work is structured as follows: in the following section a description of the different experimental procedures is presented. Next, the mesomorphic characterization of the material along with the SHG results are presented. The discussion of the results is addressed in the fourth section and finally some conclusions are drawn.

## 2. Experimental section

RM734 was synthetized following the procedure described in reference [7] and Dispersed-Orange3 was commercially available (Merck). The chemical structure of both compounds can be seen in Fig. 1a.

SHG measurements were carried out in home-made wedge cells (wedge angle = 0.0031 rad). Both glass surfaces were treated with polyvinyl alcohol. Two ITO electrodes were patterned in one of the substrates, separate 5 mm apart in order to apply in plane electric fields to the sample. Both surfaces were parallel rubbed along the electric field direction.

For the SHG measurements two different setups with optical components adapted to two different Q-switched laser sources were used ($\lambda$ = 1064 nm and $\lambda$ = 1574 nm) (See Fig.1c). Briefly, the experimental setup for 1064 nm described in detail elsewhere [32] was adapted to support the 1574 laser source. The sample was placed on a high-resolution displacement stage in order to scan the sample thickness. The beam was collimated and the spot size diameter was limited to 1 mm by a pinhole. The polarization of the fundamental light was selected with a half-wave retarder and the polarization of the SH light was checked with a polarizer (Fig. 1c). SH light was detected by a different photomultiplier for each experimental setup in order to ensure spectral sensitivity at the corresponding SH wavelength.

The characterization of the mesomorphic behaviour of the material was carried out in the same wedge sample by texture observation using polarized light microscopy (POM). We also carried out spectral absorption studies for the characterization of the SHG performance in absorbing regimen ($\lambda$ = 1064 nm). This technique was also used to determine the nematic order parameter of the mixture. For this task, homemade samples (5 μm thickness) treated for parallel homogeneous alignment were used. The sample was placed between two polarizers and illuminated by a LED source (Thorlabs

LEDD1B). The transmitted light spectrum was analyzed by means of a fiber-based spectrometer (Avantes-AvaSpec 2048) with a resolution of 0.4 nm FWHM.

In all cases, the temperature of the sample was controlled by a hot stage with 0.1ºC resolution.

3. Experimental results

Previous to the SHG studies, a mesomorphic characterization of the material was carried out by POM. The sample was cooled down from the isotropic phase down to room temperature under a weak square-wave electric field of $1 Vmm^{-1}$ and 0.5 Hz frequency. The phase transition temperature from the nematic to the FN phase can be easily determined by the onset of the ferroelectric switching in the sample. The phase sequence of the material is depicted in Fig 1a. In Fig 1b the sample texture at 110ºC is shown under the same electric field. As can be seen, the alignment is excellent in the FN phase. This is an important requirement in order to obtain a reliable characterization of the second order susceptibility tensor.

(a)

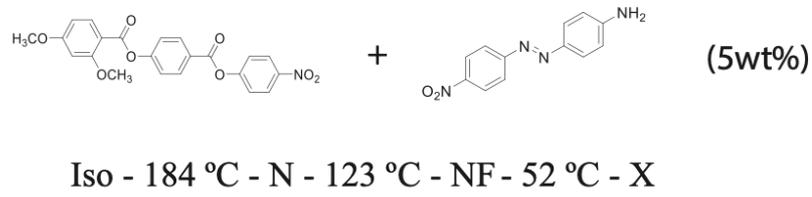

Iso - 184 ºC - N - 123 ºC - NF - 52 ºC - X

(b)

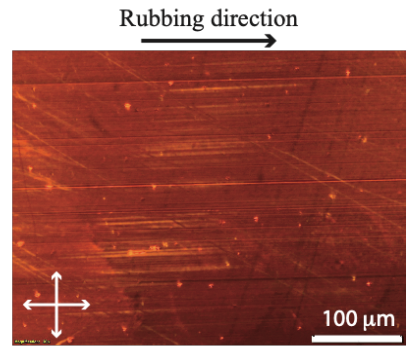

(c)

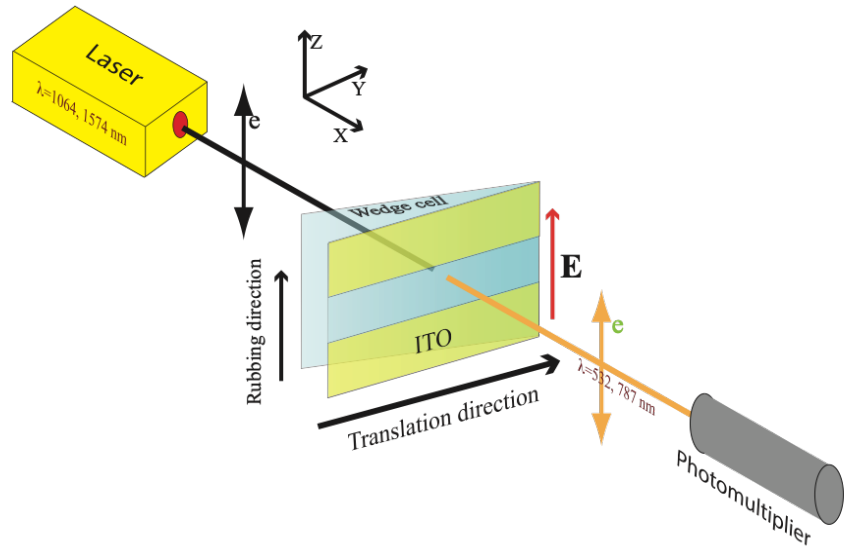

Fig 1. (a) Chemical structure of the compounds RM734 and Dispersed-Orange 3, and phase sequence of the mixture (Iso: isotropic, N: nematic, NF: ferroelectric nematic, X: crystal). (b) Texture of the mixture in the FN phase at 110ºC. White arrows indicate the polarizer directions. (c) Scheme of the experimental setups for SHG measurements at the fundamental wavelengths λ = 1064 nm and λ = 1574 nm. Double headed arrows indicate the polarization direction of the fundamental and SH lights.

The symmetry of the FN phase is ∞m, and, assuming Kleinman conditions only two different components of the $d_{ij}$ tensor appear (here we have used the Voigt two index notation). In the reference frame of Fig. 1c the most important component of the tensor is $d_{33}$. The other non-null component, $d_{13}$, is typically one order of magnitude smaller and will be neglected in this work. $d_{33}$ can be measured in the polarization configuration depicted in Fig 1c. The SHG power ($P_{2\omega}$), in this case, is given by expression [29]:

$$P_{2\omega} = AP_\omega^2 d_{33}^2 L^2 \exp\left[-\frac{\alpha_\| L}{2}\right]\frac{\sin^2\left(\frac{2\pi\Delta n L}{\lambda}\right)+\sinh^2\left(\frac{\alpha_\| L}{4}\right)}{\left(\frac{2\pi\Delta n L}{\lambda}\right)^2+\left(\frac{\alpha_\| L}{4}\right)^2} \qquad (1)$$

where $P_\omega$ and $\lambda$ are the input power and wavelength of the fundamental light, respectively, $\alpha_\|$ is the absorption coefficient for the SH light, $L$ is the sample thickness, $\Delta n = n_e(2\omega) - n_e(\omega)$ and, $A$ is a constant to be determined by calibrating the experimental setup from the maximum SH intensity of a *y*-cut quartz sample assuming the value $d_{11} = 0.4$ pmV$^{-1}$ for both fundamental wavelengths. In this expression, we neglected the absorption at the fundamental wavelength.

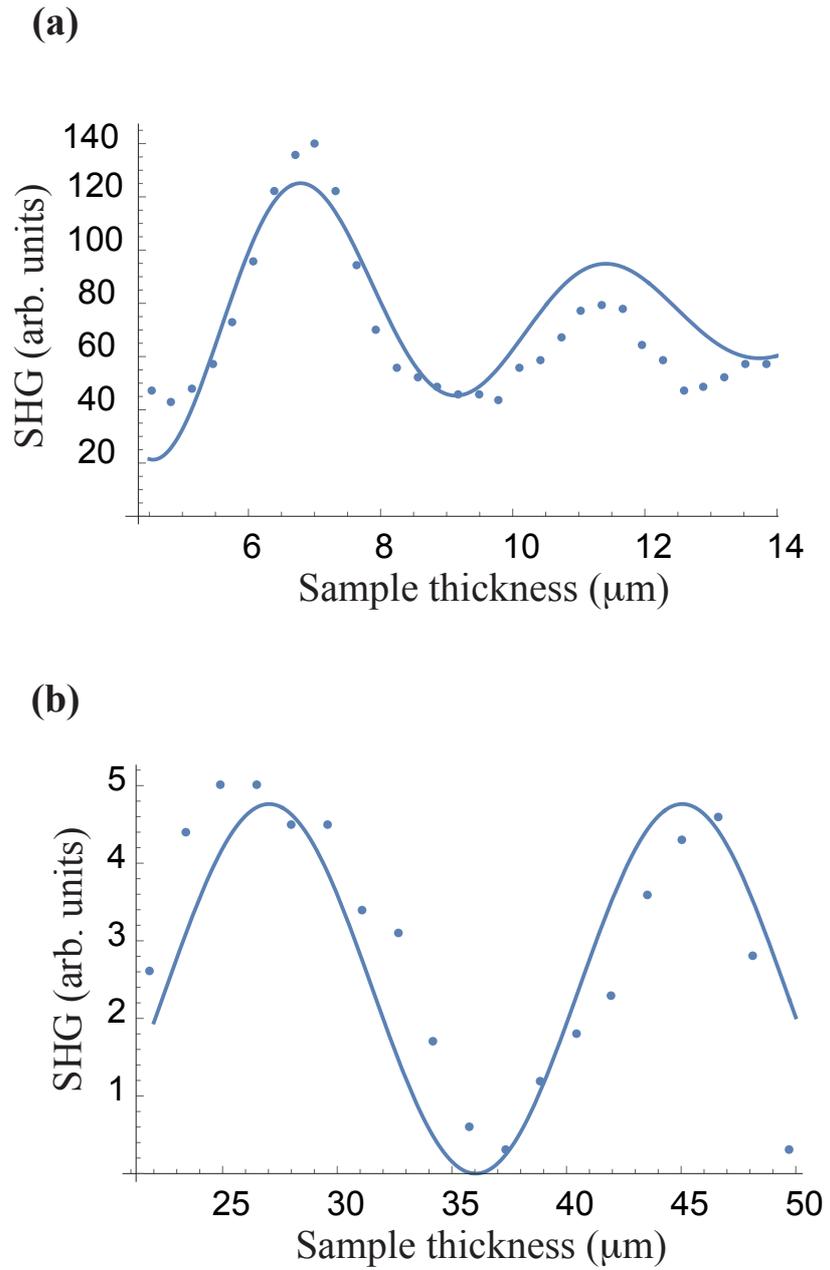

Fig. 2. SHG intensity versus sample thickness for the fundamental light wavelengths (a) 1064 nm and (b) 1574 nm. The blue continuous line is the fit to expression (1). In the first case, the absorption coefficient is $\alpha_\parallel = 0.34$ μm$^{-1}$ whereas the material is transparent for $\lambda = 1574$ nm. From each fit, two parameters are obtained, namely $\Delta n$ and $d_{33}$.

Solid circles in Fig. 2a show the measured SH power versus sample thickness at $\lambda = 1064$ nm. As can be seen, SH signal shows the typical profile of SH absorbing materials. In order to obtain the $d_{33}$ coefficient, a characterization of the absorption must first be carried out. By means of the experimental setup described above, we obtained $\alpha_\parallel = 0.34$ µm$^{-1}$. Details of the measurement will be given below. By fitting $P_{2\omega}$ versus $L$ according to expression (1) (continuous line in Fig. 2a), the following parameters are obtained: $d_{33} = 25$ pmV$^{-1}$ and $\Delta n = 0.12$. As can be seen, the experimental results fit reasonably well to expression (1). It should be notice that the fitting of the experimental data to expression (1) is subjected to important errors when the material is highly absorbent, which limits the maximum concentration to be studied. The chromophore contribution to the SHG performance is very remarkable since the pure compound efficiency is much smaller ($d_{33} = 5.6$ pmV$^{-1}$). This result is among the highest ones of liquid crystals and validates blending as good strategy for improving the SHG efficiency of FNLCs. However, the result is absorption enhanced, which means that the outstanding SHG performance cannot be fully exploited in applications.

In order to characterize the material in transparent regime, we carried out SHG measurements at $\lambda = 1574$ nm. Under the same polarization configuration, the SHG signal versus sample thickness is depicted in Fig. 2b. As can be seen, the SHG profile presents Maker fringes characteristic of a transparent material. In this case, Eq. 1 can be used with the condition $\alpha_\parallel = 0$. Continuous line represents the fit of Eq. 1 versus $L$. The obtained parameters are $d_{33} = 8.7$ pmV$^{-1}$ and $\Delta n = 0.044$. Although the decrease of the $d_{33}$ coefficient, in absence of absorption, is important, the improvement in the SHG efficient respect to the pure compound is still very remarkable. In fact, the obtained $d_{33}$ value represents a record among the liquid crystalline materials without absorption (see table 3 in reference [30]). Therefore, the suitability of blending chromophores with FNLCs is evident and should be explored exhaustively in different materials and concentrations.

In order to link the microscopic features of the nonlinear response of the material to the measured SHG, we need some additional information about the polar ($S1$) and nematic ($S2$) order parameters of the mixture. $S2$ can be obtained from absorption measurements. The absorption coefficients for incident light polarized parallel or perpendicular to the molecular director are related to $S2$ by means of expressions [33]:

$$\alpha_\parallel = C(\tfrac{1}{3} + \tfrac{2}{3} S_2) \tag{2}$$

$$\alpha_\perp = C(\tfrac{1}{3} - \tfrac{1}{3} S_2) \tag{3}$$

where $C$ is a constant. We performed absorption measurements for light at $\lambda = 532$ nm polarized parallel and perpendicular to the molecular director in the FN phase (110ºC). From the quotient of the transmitted intensity in both cases $\alpha_\parallel - \alpha_\perp = CS_2 = 0.29$ μm$^{-1}$ results. A value $C = 0.44$ μm$^{-1}$ was obtained after measuring the absorption in the isotropic phase ($S_2 = 0$). The resulting nematic order parameter $S_2 = 0.66$ is similar to that of the pure RM734 ($S_2 = 0.69$) [7], which means that, at least for a 5 wt% concentration, the incorporation of the chromophore in the material does not produce significant orientational disorder in the material.

## 4. Discussion

Now, we will connect the macroscopic SHG coefficient $d_{33}$ with the microscopic NLO features of the material. We will focus in the case of the transparent regime ($\lambda$ = 1574 nm). In the case of RM734, we will assume that the hyperpolarizability contribution is mainly due to the $NO_2$-$\pi$-O group [7]. Thus, the main donor-acceptor groups for both the RM734 and the chromophore can be considered to work parallel to the long molecular axis ($Z$ direction in the molecular frame). The corresponding second order hyperpolarizability tensor of the mixture can be approached by a single component $\beta_{333}^{eff} = (1-c)\beta_{333}^{RM734} + c\beta_{333}^{chro}$, where $\beta_{333}^{RM734}$ and $\beta_{333}^{chro}$ are the corresponding hyperpolarizability coefficients of RM734 and the chromophore, respectively, and $c$ = 0.05 is the weight concentration in the present work. The $\beta_{333}$ values for both compounds can be deduced from the literature [34]. These values where measured by means of the EFISH technique at $\lambda$ = 1907 nm and are: $\beta_{333}^{RM734}(1907)$ = $3.0 \times 10^{-30}$ esu and $\beta_{333}^{chro}(1907) = 29 \times 10^{-30}$ esu. The corresponding values at $\lambda$ = 1574 nm can be deduced by using the two-level dispersion model approach, assuming resonant wavelengths $\lambda_{max}$ = 304 nm for RM734 and $\lambda_{max}$ = 420 nm for the chromophore [34]. We obtain $\beta_{333}^{RM734}(1574) = 3.2 \times 10^{-30}$ esu and $\beta_{333}^{chro}(1574) = 33 \times 10^{-30}$ esu, leading to $\beta_{333}^{eff} = 4.7 \times 10^{-30}$ esu.

By assuming the oriented gas model, the susceptibility tensor component $d_{33}$ can be expressed as:

$$d_{33} = Nf^3 \langle \cos^3\theta \rangle \beta_{333}^{eff}, \qquad (4)$$

where $\langle\cos^3\theta\rangle$ is the thermal averaged, being $\theta$ the angle between the long molecular axis and the molecular director (Z direction in the laboratory frame), $N$ is the number of molecules per unit volume, and $f$ is the Lorentz factor $f = (n^2 + 2)/3$, being $n$ the average refractive index ($n = 1.6$). We will assume that the density of the material is similar to that of the pure RM734 compound ($\rho = 1.3$ g.cm$^{-3}$) [20], $N = 1.8 \times 10^{29}$ molecules cm$^{-3}$. In order to obtain $\langle\cos^3\theta\rangle$, we will proceed as in reference [7] and we will assume a polar order parameter equal to that reported for the pure RM734 in reference [7] ($S1 = 0.723$). The thermal average $\langle\cos^3\theta\rangle$ can be calculated provided $S1 = \langle\cos\theta\rangle$ and $S2 = \langle\frac{1}{2}(3\cos^2\theta - 1)\rangle$ are known. In our case, we obtained $\langle\cos^3\theta\rangle = 0.60$. This value is the almost same as in the pure RM734 material, where $\langle\cos^3\theta\rangle = 0.605$ [7], which means that the degree of orientational order of the chromophores is similar to that of the host molecules.

Under all these approximations, we obtain $d_{33} = 6.3$ pmV$^{-1}$. The agreement with the experimental outcome is quite reasonable, taking into account the roughness of the used approaches. It should be mentioned that the experimental value could be slightly overestimated in the calibration process since we have disregarded the dispersion of $d_{11}$ coefficient of quartz which, in fact, is expected be somewhat smaller than 0.4 pmV$^{-1}$ at $\lambda = 1574$ nm.

In the absorbing case, we cannot rely on a similar procedure to connect the macroscopic NLO coefficients to the molecular hyperpolarizabilities due to the proximity of SH light wavelength to the chromophore absorption bands, leading to an infeasible error in the calculations.

Compared to other liquid crystals designed for NLO, the SHG performance is the highest reported up to now in a transparent material. Therefore, the strategy of blending FNLC with high-performance chromophores seems to be an interesting route to explore. In addition, in our material, it can be stated that the degree of polar order induced in the chromophore molecules is similar to that of the host FNLC, which reveals the strong ability of FN molecular constituents to induced polar order in the guest compound. This feature can be exploited not only in NLO materials but can be used for the design of FN materials with different potentialities.

## 5. Conclusions

In this work, we have explored the feasibility of blending the prototype FNLC, RM734 with the strong chromophore, Dispersed Orange 3, in order to obtain a high performance SHG material. The hosting FNLC has been selected for its high SHG response even in pure state. The mixture presents stable FN phase at wide temperature range and can be easily aligned in glass cells.

One of the most outstanding results is the fact that the chromophore housed in the material presents a polar order similar to that of the pure FNLC. Therefore, the blending strategy results to be an alternative route to explore, in parallel to the design of new compounds, to obtain FNLCs aimed at different potentialities.

Regarding the SHG performance of the material, the second order susceptibility tensor has been characterized in absorbing and transparent regimens. The absorption enhanced $d_{33}$ coefficient reaches the outstanding value of 25 pmV$^{-1}$. What is even more interesting is the result in transparent regimen (8.5 pmV$^{-1}$) which results to be the highest one reported so far.

**Acknowledgements**

This work was financially supported by the Basque Government project IT1458-22, the Spanish Government project PID2023-150255NB-I00 from MCIU /AEI /10.13039/501100011033 / FEDER, the Gobierno de Aragon-FEDER research group E47_23R.

**Disclosure statement**

No potential conflict of interest was reported by the author(s).